\documentclass[prb,aps,twocolumn,floats,floatfix,superscriptaddress,longbibliography]{revtex4-2}
\usepackage{soul}
\usepackage{xcolor}
\usepackage{multirow}
\usepackage{manfnt}
\usepackage{amsmath}
\usepackage{graphicx}
\usepackage{amssymb}
\usepackage{epsfig}
\usepackage{simplewick}
\usepackage{balance}
\usepackage[colorlinks = true]{hyperref}
\begin{document}
	\newlength{\upit}\upit=0.1truein
\newcommand{\cube}{\mbox{\mancube}}
	\newcommand{\raiser}[1]{\raisebox{\upit}[0cm][0cm]{#1}}
	\newcommand{\ltappr}{{{\lower4pt\hbox{$<$} } \atop \widetilde{ \ \ \ }}}
	\newlength{\bxwidth}\bxwidth=1.5 truein
	\newcommand\frm[1]{\epsfig{file=#1,width=\bxwidth}}
	\newcommand{\cg}{{\cal G}}
	\newcommand{\dif}[2]{\frac{\delta #1}{\delta #2}}
	\newcommand{\ddif}[2]{\frac{\partial #1}{\partial #2}}
	\newcommand{\Dif}[2]{\frac{d #1}{d #2}}
	\newcommand{\str}{\hbox{Str}}
	\newcommand{\Str}{\underline{\hbox{Str}}}
	\newcommand{\tr}{{\hbox{Tr}}}
	\newcommand{\Tr}{\underline{\hbox{Tr}}}
	\newcommand{\dg}{^{\dagger }}
	\newcommand{\vk}{\mathbf k}
	\newcommand{\vq}{{\vec{q}}}
	\newcommand{\vp}{\bf{p}}
	\newcommand{\al}{\alpha}
	\newcommand{\BigL}{\biggl }
	\newcommand{\BigR}{\biggr }
	\newcommand{\gtappr}{{{\lower4pt\hbox{$>$} } \atop \widetilde{ \ \ \ }}}
	\newcommand{\si}{\sigma}
	\newcommand{\rarrow}{\rightarrow}
	\newcommand{\up}{\uparrow}
	\newcommand{\dw}{\downarrow}
	\def\fig#1#2{\includegraphics[height=#1]{#2}}
	\def\figx#1#2{\includegraphics[width=#1]{#2}}

	\newcommand{\bk}{{\bf{k}}}
	\newcommand{\bx}{{\bf{x}}}
\newcommand{\ba}{{\bf{a}}}
	\newcommand{\pmat}[1]{\begin{pmatrix}#1\end{pmatrix}}
	\newcommand{\ua}{\uparrow}
	\newcommand{\da}{\downarrow}

	\newcommand{\be}{\begin{equation}}
		\newcommand{\ben}{\begin{equation*}}
			\newcommand{\ee}{\end{equation}}
		\newcommand{\een}{\end{equation*}}
	\newcommand{\parr}{\parallel}
	\newcommand{\bmx}{\begin{array}}
		\newcommand{\emx}{\end{array}}
	\newcommand{\bean}{\begin{eqnarray*}}
		\newcommand{\eean}{\end{eqnarray*}}
	\newcommand{\dn}{^{\vphantom{\dagger}}}
	\newcommand{\nn}{\vphantom{-}}
	\newcommand{\lr}{\leftrightarrow}
	\newcommand{\ra}{\rightarrow}
	\newcommand{\la}{\leftarrow}
	\newcommand{\bb}[1]{\mathbb{#1}}
	\newcommand{\qqquad}{\qquad\qquad\qquad}
	\newcommand{\eps}{\epsilon}
	\newcommand{\sgn}[1]{{\rm sign}{#1}}
	\newcommand{\pref}[1]{(\ref{#1})}
	\newcommand{\tilda}[2]{\intopi{d{#1}}\Big(g_{#1}^A(\eps_p){#2}\Big)}
	\newcommand{\intpi}[1]{\int_{-\pi}^{+\pi}{#1}}
	\newcommand{\im}[1]{{\rm Im}\left[ #1 \right]}
	\newcommand{\trr}[1]{{\rm Tr}\Big[ #1 \Big]}

	\newcommand{\abs}[1]{\left\vert #1 \right\vert}
	\newcommand{\bra}[1]{\left\langle #1 \right\vert}
	\newcommand{\ket}[1]{\left\vert #1\right\rangle}
	\newcommand{\braket}[1]{\left\langle #1\right\rangle}
	\newcommand{\mat}[1]{\left(\bmx{cc}#1\emx\right)}
	\newcommand{\matc}[2]{\left(\bmx{#1}#2\emx\right)}
	\newcommand{\matn}[1]{\bmx{cc}#1\emx}
	\newcommand{\matl}[1]{\bmx{ll}#1\emx}
	\newcommand{\sepline}{\begin{center}\rule{8cm}{.5pt}\end{center}}

	\setlength{\parindent}{0.5cm}
	\newcommand{\indentoff}{\setlength{\parindent}{0cm}}
	\setcounter{topnumber}{9}
	\setcounter{bottomnumber}{9}
	\setcounter{totalnumber}{9}
	\renewcommand{\topfraction}{0.9}
	\renewcommand{\bottomfraction}{0.9}
	\renewcommand{\textfraction}{0.05}
	\renewcommand{\floatpagefraction}{0.5}
	\emergencystretch=3em
	\newcommand{\EJK}[1]{{\color{olive} EJK: #1}}
	\newcommand{\imEJK}[1]{{\color{olive}\bf \Large \ddag}\marginpar{\scriptsize \color{olive}\bf  \ddag #1}}
	\newcommand{\SEC}[1]{{\color{blue} \textit{#1}}}
	\newcommand{\red}[1]{{\color{red}#1}}
	\newcommand{\blue}[1]{{\color{blue}#1}}
	\newcommand\relbd{\mathrel{{\bf\smash{{\phantom- \above1pt \phantom-
	}}}}}
	\newcommand\ltdash{\raise-0.7pt\hbox{$\scriptscriptstyle |$}}
	\newlength{\figwidth}
	\figwidth=10cm
	\newlength{\shift}
	\shift=-0.2cm
	\newcommand{\fg}[3]
	{
		\begin{figure}[ht]
			\vspace*{-0cm}
			\[
			\includegraphics[width=\figwidth]{#1}
			\]
			\vskip -0.2cm
			\caption{\label{#2}
				\small#3
			}
	\end{figure}}
	\newcommand{\fgb}[3]
	{
		\begin{figure}[b]
			\vskip 0.0cm
			\begin{equation}\label{}
				\includegraphics[width=\figwidth]{#1}
			\end{equation}
			\vskip -0.2cm
			\caption{\label{#2}
				\small#3
			}
	\end{figure}}
	\graphicspath{{Figures/CPTVfigs/}}

	\title{Bose metal near pair-density-wave order in a spin-orbit-coupled Kondo lattice}
\author{Piers Coleman}
\affiliation{
Center for Materials Theory, Department of Physics and Astronomy,
Rutgers University, 136 Frelinghuysen Rd., Piscataway, NJ 08854-8019, USA}
\affiliation{Department of Physics, Royal Holloway, University
of London, Egham, Surrey TW20 0EX, UK.}

\author{Aaditya Panigrahi}
\affiliation{Department of Physics, Cornell University, Ithaca, NY 14853, USA}
\author{Alexei Tsvelik}
\affiliation{Division of Condensed Matter Physics and Materials
Science, Brookhaven National Laboratory, Upton, NY 11973-5000, USA}

	\date{\today}

\begin{abstract}
We show that a three-dimensional superconductor with a non-Abelian SU(2) order parameter can support an extended resistive regime — a Bose metal, in which transport is carried by bosonic electron-Majorana bound states — separating a uniform superconductor from a pair-density-wave (PDW) phase. The setting is a solvable Kondo lattice model introduced previously by the present authors, in which Kondo screening of a Yao-Lee $\mathbb{Z}_2$ spin liquid generates an order parameter with SU(2), rather than conventional U(1), symmetry, containing both superconducting and spin-density-wave components. Two effects cooperate to make fluctuations anomalously strong in three dimensions: the vanishing of the quadratic superconducting stiffness near the Lifshitz point where the optimal pairing momentum shifts from zero to finite $Q$, and the enlarged SU(2) order-parameter manifold. Building on our prior result that doping away from half-filling drives amplitude-modulated PDW order via finite-momentum electron-Majorana condensation, we analyze the fluctuation-dominated regime above that phase using a nonlinear sigma model. We find that the order-parameter propagator develops a ring of soft modes throughout the disordered phase, and that at the smallest temperatures the resulting resistivity  $R \sim T^3$ in three dimensions.
\end{abstract}
	\maketitle

\section{Introduction}

Pair-density-wave (PDW) order --- spatially modulated superconductivity --- has been observed in a growing class of unconventional superconductors~\cite{agterberg_physics_2020}, yet the nature of the transition into the PDW state, particularly in three dimensions, remains poorly understood. A basic question is whether the system passes directly from uniform superconductivity into PDW order, or whether fluctuations can sustain a resistive intermediate state. Here we show that when the order parameter has SU(2) rather than conventional U(1) symmetry, the latter is possible: a broad Bose-metal regime, carrying charge through bosonic electron-Majorana bound states, emerges between the two ordered phases. Two effects cooperate to make this happen.

The first is the vanishing of the quadratic superconducting stiffness, which softens the order-parameter fluctuations. Since the transition from uniform superconductivity to a PDW is accompanied by a shift of the wave vector at which the pairing susceptibility is maximal, this softening naturally occurs near a Lifshitz transition. The second is the size of the symmetry manifold of the order parameter: the larger the manifold, the stronger the fluctuations. In the microscopic system studied here, the order parameter lives on SU(2) rather than the conventional U(1), making both effects unusually pronounced.

This question is nontrivial because a PDW requires a pairing susceptibility peaked at finite momentum, something that does not arise naturally in a BCS superconductor~\cite{BCS_Theory}. In the BCS problem, the logarithmic divergence of the uniform susceptibility is cut off by the finite momentum $Q$---more precisely by $v_FQ$, where $v_F$ is the Fermi velocity. In conventional FFLO physics, a Zeeman field compensates for this by polarizing the Fermi surfaces and favoring pairing at nonzero momentum~\cite{fulde_superconductivity_1964,larkin_nonuniform_1964}. By contrast, experiments suggest that PDW order can develop spontaneously in unconventional superconductors even without an applied magnetic field, and the origin of this tendency remains unsettled.

The route we take is guided in part by ideas developed in the context of quark-gluon plasma, where a related fluctuation mechanism suppresses order-parameter condensation~\cite{QG1,QG2}. An Abelian version of the Lifshitz problem was analyzed in Ref.~\cite{Andrii}; here we extend the discussion to a genuinely non-Abelian order parameter.

The concrete microscopic setting is a solvable three-dimensional model introduced previously by the present authors (CPT, for Coleman-Panigrahi-Tsvelik), in which Kondo screening of a $\mathbb{Z}_2$ spin liquid induces order in the surrounding conduction sea~\cite{coleman_solvable_2022,tsvelik_order_2022}. The resulting order parameter has SU(2) symmetry and contains both superconducting and spin-density-wave components. The underlying spin liquid is the Yao-Lee state, a three-dimensional generalization of the spin-orbital $\mathbb{Z}_2$ Kitaev spin liquid~\cite{yao_fermionic_2011}. The model remains tractable because the single-occupancy constraint is built into the Majorana representation rather than imposed through a separate mean-field procedure.

At half-filling, the electron, hole, and Majorana Fermi surfaces are nested, producing a logarithmic instability to electron-Majorana pair condensation under arbitrarily weak Kondo coupling and hence odd-frequency triplet superconductivity. In our previous work~\cite{CPT2025}, we showed that doping away from half-filling shifts the electron and hole Fermi surfaces while leaving the Majorana Fermi surface unchanged. That mismatch moves the preferred pairing channel to finite momentum, favoring an FFLO-like electron-Majorana condensate and producing amplitude-modulated PDWs.

The present paper focuses on the fluctuation regime near that transition. We first formulate a Ginzburg-Landau theory for the SU(2) order parameter and use it to identify the soft sector near the Lifshitz point. We then turn to a nonlinear sigma model, following the strategy of Refs.~\cite{QG1,QG2}, to analyze the fluctuation-dominated regime itself. Finally, we estimate the corresponding fluctuation conductivity and discuss what it implies for PDW formation in the CPT model. Details of the microscopic derivation are collected in the Appendix.

\section{Ginzburg-Landau theory for a non-Abelian order parameter\label{sec:GL}}

We now translate this physical picture into an effective field theory. The microscopic derivation is deferred to the Appendix; here we isolate the ingredients needed for the fluctuation analysis. The natural starting point is a Ginzburg-Landau description of the SU(2)-symmetric order parameter, with free-energy density
\begin{equation}
\begin{aligned}
F ={}& \sum_q \frac{1}{2}{\cal V}(q)\dg\bigl[1/J_K -\chi(q;T,\mu)\bigr]{\cal V}(q) \\
&+ \int d^3 x \frac{g}{2}({\cal V}\dg{\cal V})^2,
\end{aligned}
\label{GL1}
\end{equation}
where ${\cal V}$ is a two-component spinor transforming in the spin-$\tfrac{1}{2}$ representation of SU(2), $\chi$ is the pairing susceptibility, and $J_K$ is the Kondo coupling. In the CPT model, ${\cal V}$ is a bound state of an electron (charge $e$, spin $\tfrac{1}{2}$) and a Majorana fermion (charge zero, spin $1$). Strictly speaking, ${\cal V}$ is not gauge invariant because it carries both $\mathbb{Z}_2$ and U(1) charges, so the corresponding gauge fields should be included. This can in principle lead to confinement. Nevertheless, calculations for the XY version of the model~\cite{KuklovCT} indicate a very large confinement length, and the gauge-invariant order parameter still inherits the non-Abelian character of its constituents. It can be represented by three mutually perpendicular vectors,
\begin{eqnarray}
 &&{\bf d}_1 = {\cal V}\dg\vec\sigma{\cal V}, \\
 &&{\bf d}_2 +i{\bf d}_3 = i{\cal V}^T\sigma^y\vec\sigma{\cal V}, ~~{\bf d}_2 -i{\bf d}_3 = i{\cal V}\dg\sigma^y\vec\sigma{\cal V}^*,  \nonumber
\end{eqnarray}
whose non-Abelian structure is central to the fluctuation physics discussed below.

To connect this order parameter to the onset of PDW order, we expand the pairing susceptibility in momentum,
\begin{equation}
  \chi(q;T,\mu) = \chi(0,T,\mu) - \rho(T,\mu)q^2 + O(q^4).
\end{equation}
The cubic symmetry of the lattice constrains the quartic term to the form
\begin{equation}
\begin{aligned}
\chi(q)/\chi(0) -1 ={}& - A(q_x^2+q_y^2 +q_z^2) \\
&+ B(q_x^2+q_y^2 +q_z^2)^2 \\
&+ C(q_x^2q_y^2 +q_y^2q_z^2 +q_x^2q_z^2).
\end{aligned}
\label{cubic}
\end{equation}
Neglecting the cubic anisotropy ($C=0$) for the moment and retaining only the leading terms in this expansion, we arrive at the real-space Ginzburg-Landau functional
 \begin{eqnarray}
 &&F = \int d^3 x \Big[\frac{\rho}{2}(\nabla{\cal V}\dg\nabla{\cal V}) + \frac{\rho_2}{2}(\nabla^2{\cal V}\dg\nabla^2{\cal V}) + \nonumber\\
 && \frac{\tau}{2}({\cal V}\dg{\cal V}) + \frac{g}{4}({\cal V}\dg{\cal V})^2\Big], \label{GL2}
\end{eqnarray}
where $\tau = 1/J_K - \chi(0;T,\mu)$.

\section{Nonlinear sigma model and fluctuation-dominated regime\label{sec:nlsm}}

\begin{figure}[t]
            \centering
            \includegraphics[width=0.9\linewidth]{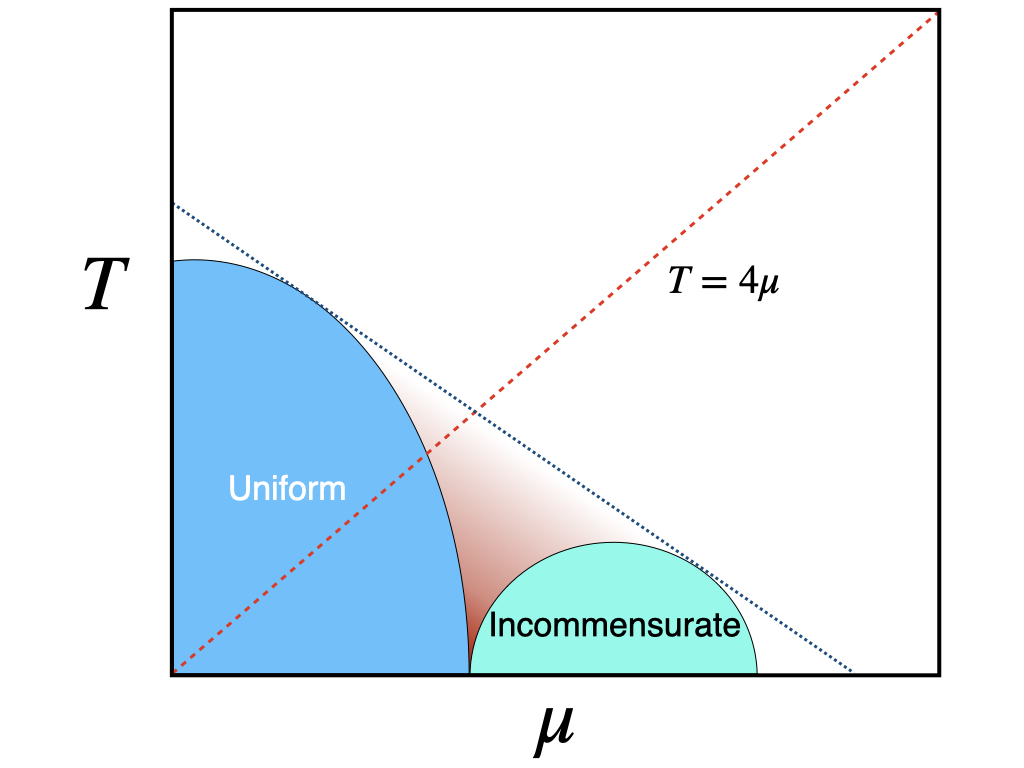}
            \caption{Suggested phase diagram. The uniform and incommensurate PDW phases approach one another near a possible quantum critical point. There is a disordered (resistive) region in which the order-parameter propagator has a maximum on a ring of wave vectors. The blue dashed line marks the onset of the single-particle spectral gap, given by Eq.~(\ref{Tc}); the orange dashed line marks where $\chi''(q)$ changes sign, which for $t=K$ occurs at $T=\mu/4$. The boundary of the IC phase is determined by condition~(\ref{IC}), where order-parameter fluctuations are suppressed by Fermi-surface anisotropy.
            }
            \label{fig:Phase_D}
        \end{figure}

The Ginzburg-Landau functional already shows that the phase structure is controlled by two temperature scales. The first, $T_{c1}$, is defined by $\tau(T_{c1},\mu)=0$, and the second, $T_{c2}$, by $\rho(T_{c2},\mu)=0$. If $T_{c2}>T_{c1}$, the transition proceeds directly into the incommensurate (IC) phase. Otherwise, the system first enters the  uniform phase, and the pairing susceptibility must be recomputed to determine whether a second transition occurs at lower temperature. The resulting schematic phase diagram is shown in Fig.~\ref{fig:Phase_D}.

This diagram also makes clear why mean-field theory fails near the point where the two scales approach one another. As $T_{c1}$ approaches $T_{c2}$, the stiffness $\rho$ vanishes and the propagator of ${\cal V}$ takes the soft form
  \begin{equation}
     \langle\langle{\cal V}(q){\cal V}\dg(q)\rangle\rangle \sim (\rho q^2 + \rho_2 q^4 + \tau)^{-1}.
  \end{equation}
When $\rho\leq 0$, thermal fluctuations diverge in the isotropic problem even in three dimensions, pushing the transition temperature to zero. Lattice anisotropy restores a finite transition temperature, but for weak anisotropy the fluctuation corrections remain substantial.

To analyze this regime more systematically, we assume that the transition temperature lies well below the quasiparticle gap and treat $\rho$ and $\rho_2$ phenomenologically. Completing the square in Eq.~(\ref{GL2}),
\begin{equation}
\frac{\tau}{2}({\cal V}\dg{\cal V}) + \frac{g}{4}({\cal V}\dg{\cal V})^2 = \frac{g}{4}\Big[({\cal V}\dg{\cal V})+ \tau/g\Big]^2 - \tau^2/4g,
\end{equation}
and substituting
\begin{equation}
{\cal V}_{\sigma} = |V_{MF}|z_{\sigma}, \qquad \sum_{\sigma}z\dg_{\sigma}z_{\sigma} =1,
\end{equation}
where $V_{MF} = (-\tau/g)^{1/2}$, integrate out the massive amplitude mode and obtain the nonlinear sigma model
\begin{eqnarray}
&& F  = \int d^3x\Big\{\frac{\tilde\rho}{2}(\nabla z\dg\nabla z) + \frac{\tilde\rho_2}{2}(\nabla^2 z\dg)(\nabla^2 z)\Big\},\nonumber\\
&&\sum_{\sigma}z\dg_{\sigma}z_{\sigma} =1, \label{PCF}
\end{eqnarray}
where $(\tilde\rho, \tilde\rho_2) = |V_{MF}|^2(\rho,\rho_2)$.

This form makes the role of fluctuations explicit. Exactly at $\rho=0$ there is no long-range order. For negative $\rho$, the would-be divergence is cut off either by the lattice anisotropy $C\neq 0$ {(see Fig. 4 as an illustration) or by quantum effects, since the integral $\int d\omega\,d^3q$ converges. This in turn opens up the possibility of a quantum critical point separating the commensurate and incommensurate phases.

To obtain an analytic description of the disordered regime, we replace the local constraint in~(\ref{PCF}) by a global one and obtain the propagator
  \begin{equation}
  \langle\langle {\cal V}(q){\cal V}\dg(q)\rangle\rangle = [\rho_2(q^2 - Q^2)^2 + m^2]^{-1}, \label{prop}
\end{equation}
where $m$ is determined self-consistently. Neglecting anisotropy, the propagator~(\ref{prop}) provides a simple description throughout the disordered phase, which remains resistive and therefore has finite conductivity. Imposing the self-consistency condition ${\cal V}{\cal V}\dg = |V_{MF}|^2$ then gives
\begin{eqnarray}
&& \frac{T}{(2\pi)^3}\int \frac{d^3 q }{\rho_2(q^2 - Q^2)^2 +m^2} = |V_{MF}|^2,\nonumber\\
&& 8\pi\rho_2 |V_{MF}|^2= (T\rho_2^{1/2}/m)\Re e \sqrt{Q^2+i m/\rho_2^{1/2}} \label{mass}
\end{eqnarray}
This yields a finite correlation length throughout the entire disordered phase separating the commensurate and incommensurate ordered phases. The solution is 
\begin{eqnarray}
 && m^2(T) = \rho_2 (T/T_0)^2\Big[(T/T_0)^2 + 2Q^2\Big], \nonumber\\
 && T_0= 8\sqrt 2 \rho_2|V_{MF}|^2.  \label{corr} 
\end{eqnarray}

Long-range order is recovered only once lattice anisotropy regularizes the problem. When $C \neq 0$, the integral of the fluctuation propagator over $q$ becomes convergent. For $|C| \ll B$, anisotropy becomes the dominant cutoff when $|C|\rho Q^4 \gtrsim m^2$, i.e., when
\begin{equation}
\tilde T < (|C|Q^2)^{1/2}. \label{IC}
\end{equation}
The extent of the fluctuation regime is therefore controlled by two quantities: the Fermi-surface anisotropy and the magnitude of the PDW wave vector. When the anisotropy is not too large, these two scales allow a broad disordered region of strong superconducting fluctuations to separate the commensurate and incommensurate phases.

\section{Fluctuation conductivity\label{sec:cond}}

 Once the fluctuating regime has been identified, the next question is how it manifests in transport, in particular in electric resistivity. We suggest that we can follow the logic of the 2D XY model  above the Berezinskii-Thouless-Kosterlitz (BKT) transition. As in that case the quasiparticle excitations are gapped and the current is carried by  a charged bosonic field. Above the transition (which in our case occurs at $T=0$) this field is not condensed and there is a finite correlation length 
\begin{eqnarray}
 && \xi^{-1}(T) = \Big(\frac{\sqrt{(m/\rho_2)^2 +Q^4}-1}{\sqrt{(m/\rho_2)^2 +Q^4}+1}\Big)^{1/2} (Q^4 + (m/\rho_2)^2)^{1/4}  =\nonumber\\
 && \Big[\frac{T^2}{T^2 +2T_0^2Q^2}\Big]^{1/2}\Big[(T/T_0)^2 +2Q^2\Big]^{1/2}.
\end{eqnarray} 
Halperin and Nelson \cite{halperin_nelson_1979} calculated the resistivity using the semiclassical approach; they argued that the finite correlation length means that there is a residual amount of normal phase where all the dissipation occurs. 

 The dimension considerations then yield the following estimate for the resistivity:
 \begin{equation}
     R(T) \sim \rho_n \xi^{-3}(T),
 \end{equation}
 where $\rho_n$ is the normal state resistivity.
 

 
According to (\ref{corr}) $R(T) \sim T^3$ throughout the entire fluctuational region. The prefactor includes the resistivity in the normal state which comes from the processes not included in our model.

\section{Conclusions\label{sec:concl}}

We have shown that PDW order with non-Abelian symmetry can support unusually strong thermal fluctuations even in three spatial dimensions, where one ordinarily expects mean-field theory to work well. The central reason is the cooperation of two effects: the softening of the order-parameter propagator as the quadratic stiffness $\rho$ changes sign near the transition between uniform superconductivity and the PDW, and the enlarged SU(2) order-parameter manifold.

In the fluctuation-dominated regime above the PDW transition, the system enters a resistive state in which transport is carried by bosonic bound states of electrons and Majorana fermions. The breadth of the fluctuation regime is controlled by the PDW wave vector and the Fermi-surface anisotropy. 

Although our analysis is grounded in the specific CPT Kondo lattice model with a Yao-Lee spin liquid, the underlying mechanism is more general. In particular, the same combination of soft modes and strong fluctuations has a close analogue in quark-gluon plasma~\cite{QG1,QG2}, where condensation is suppressed by related physics.

These ideas may be relevant to heavy-fermion systems such as UTe$_2$, where PDW order and fluctuation-driven phenomena have been discussed in the experimental and theoretical literature~\cite{gu_detection_2023,aishwarya_melting_2024}. More broadly, the electron-Majorana spinor order parameter suggests how superconductivity emerging from fractionalized degrees of freedom may broaden the landscape of possible broken-symmetry states, with possible extensions to topological phases discussed in related work~\cite{zhuang_topological_2024,panigrahi_breakdown_2024}.

\begin{acknowledgments}
We would like to thank Subir Sachdev for fruitful discussions. This work was supported by the Office of Basic Energy Sciences, Materials Sciences and Engineering Division, U.S. Department of Energy (DOE) under Contracts No.\ DE-SC0012704 (AMT) and DE-FG02-99ER45790 (PC). AP was supported by
the U.S. Department of Energy, Office of Science, Basic Energy Sciences, Materials Sciences and Engineering
Division.
\end{acknowledgments}

\appendix

\section{Microscopic model and pairing susceptibility\label{sec:model}}

\begin{figure}[!htbp]
        \centering
        \includegraphics[width=1.\linewidth]{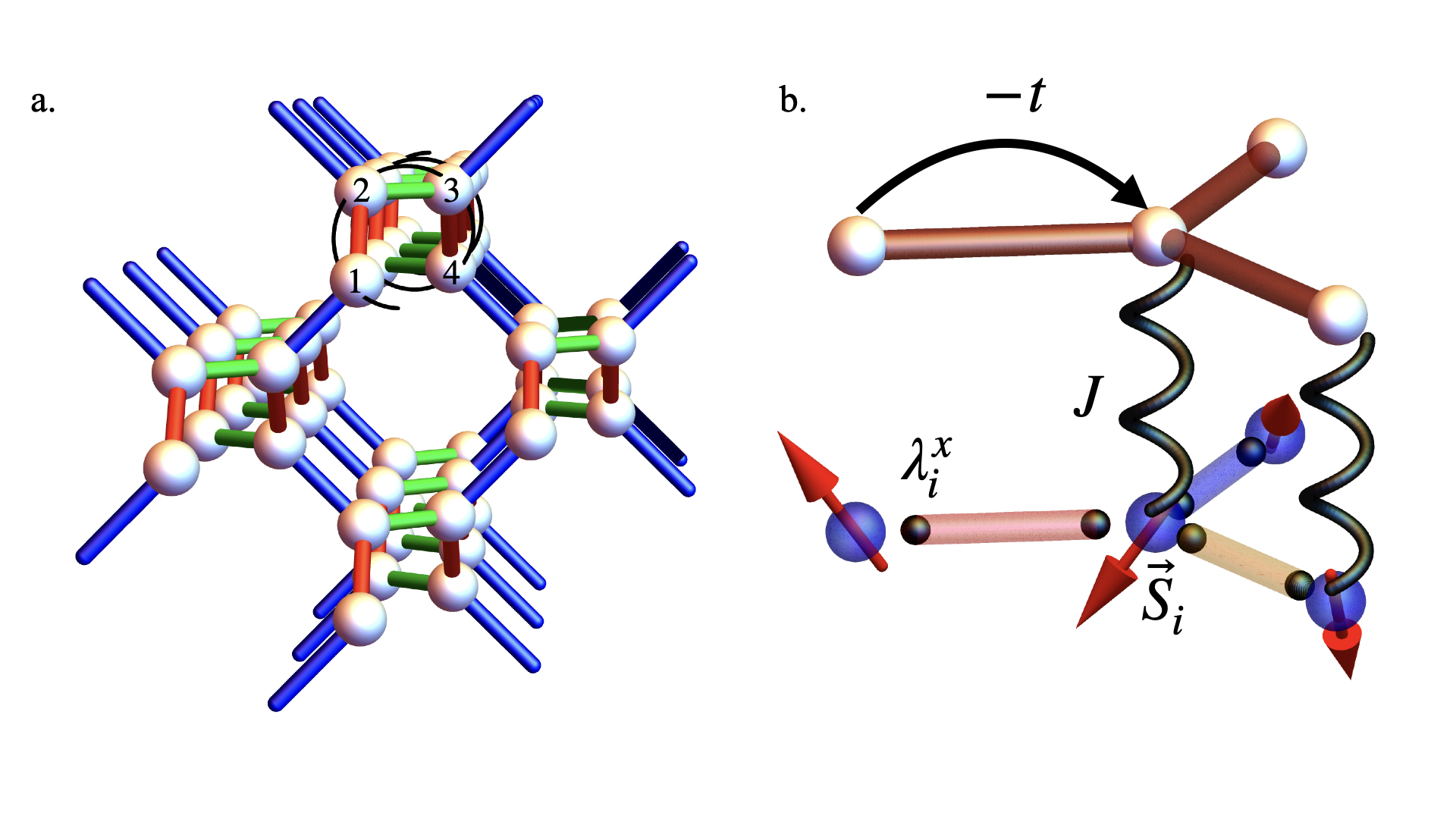}
        \caption{(a) The hyperoctagon lattice, which has a body-centered cubic (BCC) structure formed by interpenetrating square and octagonal spirals, with a four-atom unit cell~\cite{hermanns_quantum_2014}. The frustrated orbital interactions are denoted by color-coded red, blue, and green bonds. (b) In the CPT model each site is trivalent, hosting conduction electrons (linked by golden bonds), localized spins, and localized orbitals (linked by color-coded orbital interactions). The frustrated orbital interactions induce a Majorana fractionalization of the spins, which are then Kondo-coupled to the conduction electrons on the same lattice. }
        \label{fig:Hyperoctagon}
    \end{figure}

We now summarize the microscopic ingredients underlying the effective theory used in the main text. The CPT model~\cite{coleman_solvable_2022} is defined on the hyperoctagon lattice (Fig.~\ref{fig:Hyperoctagon}) and Kondo-couples conduction electrons to the spins of a Yao-Lee spin liquid. Each site possesses three degrees of freedom --- conduction electrons, localized spins, and localized orbitals --- so the Hamiltonian naturally decomposes into three components,
        $H_{CPT}=H_c+H_{YL}+H_K,$
				\label{CPTHam}
where $H_c$ describes nearest-neighbor hopping of the conduction electrons, $H_{YL}$ captures the Yao-Lee spin-spin interaction, and $H_K$ couples the conduction sea to the Yao-Lee spin liquid through the Kondo interaction:
        \begin{align}
			H_{c}=&-t\sum_{\braket{i,j}} ( c\dg_{i\si}c_{j\si}+{\rm H.c.})-\mu\sum_{i}c\dg_{i\si}c_{i\si},\nonumber
			\\H_{YL}=&K/2\sum_{\braket{i,j}}\lambda^{\al_{ij}}_i\lambda^{\al_{ij}}_j
			(\vec{S}_i\cdot\vec{S}_j),\label{YLH}
			\\H_{K}=&J\sum_{i}(c\dg_{i}\vec{\si}c_{i})\cdot \vec{S}_{i}.
			\label{CPTComp} \nonumber
        \end{align}
The Yao-Lee term is a nearest-neighbor Ising interaction in the orbital components $\lambda_j^{\alpha_{ij}}$ ($\alpha_{ij} = x, y, z$), decorated by a Heisenberg exchange between the spins $\vec{S}_j$. The orbital frustration in $H_{YL}$ induces large quantum fluctuations, leading to a fractionalization of the spins and orbitals into Majorana fermions,
\begin{equation}
\vec{\lambda}_j = -i \vec{b}_j \times \vec{b}_j, \qquad \vec{S}_j = -\frac{i}{2} \vec{\chi}_j \times \vec{\chi}_j,
\end{equation}
where $b$ and $\chi$ are Majorana fermions. At low energies, the orbital degrees of freedom decouple as static $\mathbb{Z}_2$ gauge fields that freeze into a flux-free configuration at low temperatures. The low-energy physics of the Yao-Lee spin liquid is then described by a single-band Hamiltonian
  \begin{equation}\label{MajoranaHam}
    H_{YL}=\sum_{{\bf k}\in \cube} \epsilon_{\chi}({\bf k})  \vec{\chi}^{\dg}_{{\bf k}} \cdot \vec{\chi}_{{\bf k}},
\end{equation}
where the momentum sum runs over the cubic Majorana Brillouin zone $\cube$ and the dispersion is $\epsilon_{\chi}({\bf k})=K\tilde{\epsilon}({\bf k})$, with
  \begin{equation} \label{ProjectedBand2}
 \tilde{\epsilon}({\bf k})=\frac{1}{2  Si_{x}Si_{y}Si_{z}} \left[(C^2_x +C^2_y +C^2_z)-\frac{3}{4}\right],
\end{equation}
and $C_i=\cos(k_i/2)$, $Si_i=\sin(k_i/2)$.

The low-energy physics of the conduction band~\cite{coleman_solvable_2022} is likewise described by a single band $\epsilon_c({\bf k})=-t \tilde{\epsilon}({\bf k})$, with $\tilde{\epsilon}({\bf k})$ as in Eq.~(\ref{ProjectedBand2}). Rewriting the conduction Hamiltonian in terms of the Balian-Werthamer spinor $\psi_{\bf k}=(c_{{\bf k}}, -i \sigma^2 c^{\dg}_{-{\bf k}})^T$, we obtain
\begin{equation} \label{BLHcond}
    H_{c}=\sum_{{\bf k}\in \cube}\psi^{\dagger}_{{\bf k}}(\epsilon_{c}({\bf k})\mathbb{I}-\mu  \tau_3)\psi_{{\bf k}}.
\end{equation}
Once the low-energy degrees of freedom have been identified, the Majorana fractionalization of the spins allows the Kondo interaction to be rewritten as
	\begin{equation}\label{KondoInteraction}
		H_{K}=
 - \frac{J}{2}\sum_{l}c\dg_{j}
(\vec{\chi }\cdot \vec{ \sigma })^{2}c_{j}.
	\end{equation}
This interaction is decoupled via a Hubbard-Stratonovich transformation. Introducing the self-consistent spinor field $\mathcal{V}_j$, with components $V_j = -J\langle\vec{\chi}_j\cdot\vec{\sigma}\,c_j\rangle$, gives the mean-field interaction:
    \begin{equation}\label{MFTHam}
    H_{int}[j]=\frac{1}{2}\left(\psi^{\dg}_j (\vec{\si}\cdot \vec{\chi}_j) \mathcal{V}_j+\mathcal{V}^{\dg}_{j}(\vec{\si}\cdot\vec{\chi}_j)\psi_j \right)+\frac{\mathcal{V}^{\dg}_j \mathcal{V}_j}{J},
\end{equation}
where $\psi_{\vec{k}} = (c_{\vec{k}}, -i \sigma^2 c^{\dagger}_{-\vec{k}})^T$ and $\mathcal{V}_j = (V_j, -i \sigma^2 V^*_j)^T$. In previous work~\cite{coleman_solvable_2022}, we showed that at half-filling the model exhibits a logarithmic singularity in the electron-Majorana pair susceptibility, leading to odd-frequency triplet superconductivity.

{\it Pair-density-wave instability.}
With this structure in hand, the mechanism for PDW formation follows naturally~\cite{CPT2025}. The chemical potential plays a role analogous to a Zeeman field in a conventional FFLO problem: it splits the electron and hole Fermi surfaces while leaving the Majorana Fermi surface unchanged [Fig.~\ref{fig:Fermi_Surface}(a)]. As a result, the fractionalized spinor order parameter $\mathcal{V}_{\vec q}$ acquires a finite ordering wave vector [Fig.~\ref{fig:Fermi_Surface}(b)].

        \begin{figure}[t]
            \centering
            \includegraphics[width=0.95\linewidth]{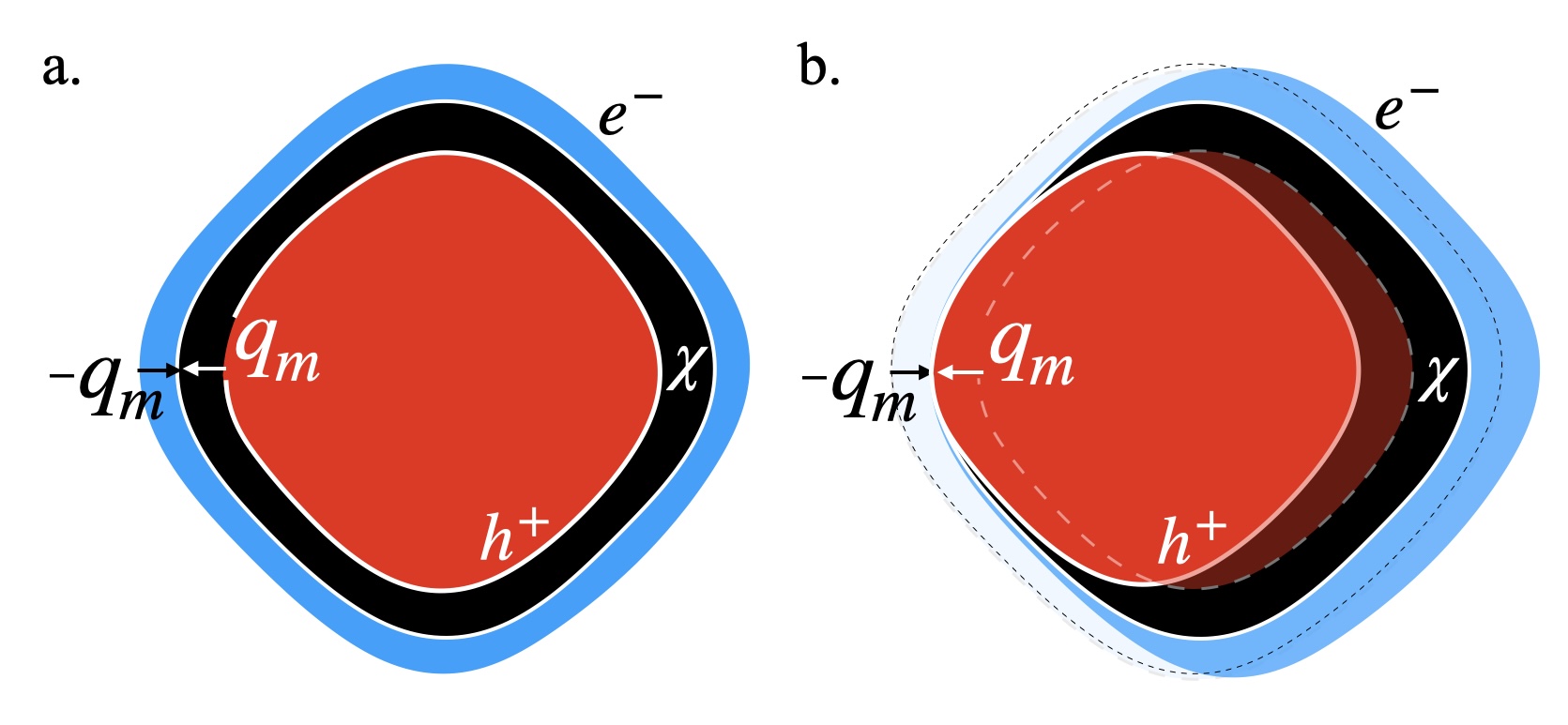}
            \caption{Impact of electron doping on the conduction sea within the hyperoctagon lattice. (a)~The conduction sea consists of electron (blue) and hole (red) Fermi surfaces, which expand and contract, respectively, as electron doping shifts the system away from half-filling. The Majorana Fermi surface (black) is unaffected by doping. (b)~Doping stabilizes finite-momentum configurations of the electron-Majorana spinor order by shifting the electron and hole Fermi surfaces by $\pm\vec{q}_m$ to nest with the Majorana Fermi surface. (the figure is taken from Ref.\cite{CPT2025})
            }
            \label{fig:Fermi_Surface}
        \end{figure}

The same physics is encoded in the static electron-Majorana pairing susceptibility, which is maximized at a finite momentum $\vec{q}$ (Fig.~\ref{fig:Susceptibility1}):
        \begin{eqnarray}\label{Susceptibility1}
    &&\frac{\partial^2 F}{\partial V^2}=\chi_{v}(\vec{q})=\\
    && -\frac{T}{4}\sum_{i\omega_n,k\in \cube}\Tr\left[G_{\psi}(\vec{k}+\vec{q},i\omega_n)\sigma^a \mathcal{Z}\mathcal G_{\chi^a}(\vec{k},i \omega_n)\mathcal{Z}^{\dagger}\sigma^a\right],\nonumber
        \end{eqnarray}
expressed in terms of the conduction and Majorana Green's functions $G_{\psi}(\vec{k}+\vec{q},i\omega_n)$ and $G_{\chi}(\vec{k},i\omega_n)$, with $\mathcal{Z}=\frac{2\mathcal{V}}{V}$ denoting the orientation of the spinor order. After straightforward simplification, the susceptibility for the hyperoctagon lattice can be written in closed form:
\begin{equation}\label{HyperoctagonSusceptibilty2}
\begin{aligned}
\chi_v (\vec{q})={}&\frac{3\rho(0)}{K+t}\log\left(\frac{(K+t) D}{\mu}\right) \\
&-\frac{3\rho(0)}{4(K+t)}\int \frac{d\Omega}{4\pi}
\Bigg[\log\left\vert 1-\frac{\vec{v}_F(\theta,\phi)\cdot \vec{q}}{\mu}\right\vert \\
&\hspace{5.7em}+\log\left\vert \frac{K}{t}+\frac{\vec{v}_F(\theta,\phi)\cdot \vec{q}}{\mu}\right\vert\Bigg].
\end{aligned}
\end{equation}
Here $\rho(0)$ is the density of states at the Fermi surface and $\vec{v}_F=\vec{\nabla}_{\bf k} \epsilon_{c}({\bf k})$ is the Fermi velocity of the conduction sea, which depends on the orientation $(\theta,\phi)$ for the hyperoctagon lattice.

       \begin{figure}[!htbp]
            \centering
            \includegraphics[width=0.8\linewidth]{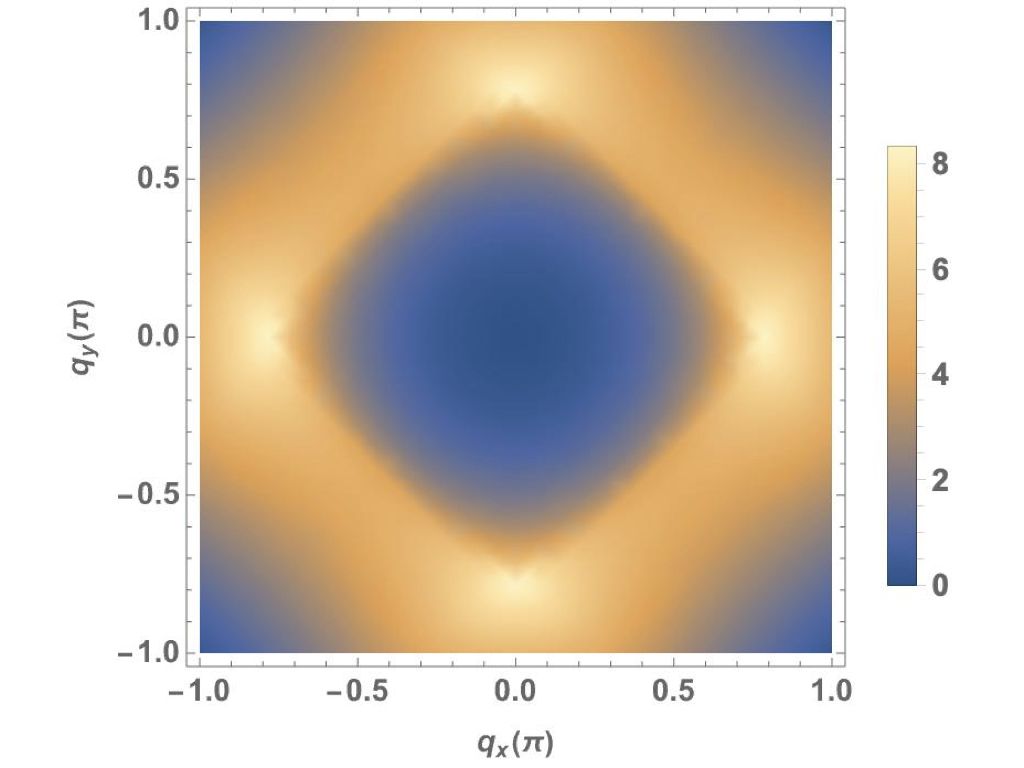}
            \caption{Finite-momentum susceptibility $\chi_v(\vec{q})$ for electron-Majorana spinor order along the $q_z = 0$ plane. The maximum occurs at $\vec{q}_i = q_m(\pm 1, 0, 0)$ and $q_m(0, \pm 1, 0)$ with $q_m = 0.7738\pi\mu/t$, showing four-fold degeneracy. Cubic symmetry yields six equivalent maxima along $\pm\hat{x}$, $\pm\hat{y}$, $\pm\hat{z}$ (the figure is taken from Ref.\cite{CPT2025}).
 }
            \label{fig:Susceptibility1}
        \end{figure}

Equation~(\ref{HyperoctagonSusceptibilty2}) makes the origin of the PDW instability explicit. The logarithmic divergence of the electron-Majorana susceptibility is cut off by the chemical potential, which splits the nesting between the conduction sea and the Majorana Fermi surfaces. The result is a set of susceptibility maxima at finite momenta $\vec{q}$, shown in Fig.~\ref{fig:Susceptibility1}.

To locate these maxima more explicitly, we express the Fermi velocity of the conduction sea in angular coordinates,
\begin{equation}\label{FermiVelocity2}
    \vec{v}_F=\frac{1}{2 s_x s_y s_z}\left(\frac{c_x^3}{s_x},\frac{c_y^3}{s_y},\frac{c_z^3}{s_z}\right),
\end{equation}
and parameterize the $c_i$ on the Fermi surface in spherical coordinates,
\begin{equation}\label{angCos}
    (c_x,c_y,c_z)=(r\cos\phi\sin\theta,\, r\sin\phi\sin\theta,\, r\cos\theta),
\end{equation}
in terms of angular variables $\theta$ and $\phi$ and radius $r=\sqrt{3}/2$. Carrying out the angular integral in Eq.~(\ref{HyperoctagonSusceptibilty2}) numerically, we find that the susceptibility is maximized (Fig.~\ref{fig:Susceptibility1}) at
\begin{equation}
\vec{q}_i \in \{q_m(\pm 1,0,0),\, q_m(0,\pm 1,0),\, q_m(0,0,\pm 1)\},
\label{maxima_vectors}
\end{equation}
along the principal axes, where the wavenumber at the susceptibility maximum is $q_m = 0.77\pi \mu/t$, proportional to the chemical potential. These results are consistent with the general symmetry considerations of Eq.~(\ref{cubic}): the cubic symmetry of the model yields six degenerate wave vectors along which the spinor order prefers to modulate. Such degeneracy is often lifted, as in FFLO states~\cite{larkin_nonuniform_1964}, where attractive interactions between $\mathcal{V}_{\vec{q}}$ and $\mathcal{V}_{-\vec{q}}$ favor amplitude modulation of the form $\mathcal{V}_{\vec{q}} + \mathcal{V}_{-\vec{q}}$, or other superpositions depending on the level of doping.

\section{Microscopic stiffness and phase diagram\label{sec:stiff}}

Having established the finite-momentum instability microscopically, we now extract the stiffness parameters that control the effective theory. The subsequent calculations are performed for the CPT model described in Appendix~\ref{sec:model}. As noted in the Introduction, the chemical potential plays a role analogous to the Zeeman field in the FFLO scenario. It has two effects: it lowers the transition temperature $T_c$, and it reduces the quadratic coefficient $\rho(T,\mu)$ (stiffness) in the pairing susceptibility, eventually changing its sign.

Taking Eq.~(\ref{Susceptibility1}) and, to simplify the calculations, considering the case of equal dispersions ($t=K$), the pairing susceptibility in the normal state is
        \begin{equation}
        \begin{aligned}
        \chi(q) ={}& \frac{3}{4}\int \frac{d^3 k}{(2\pi)^3}
        \Bigg\{\frac{\tanh\!\left[\frac{\epsilon(k+q/2)+\mu}{2T}\right]
        + \tanh\!\left[\frac{\epsilon(k-q/2)}{2T}\right]}{\epsilon(k+q/2)+\epsilon(k-q/2)+\mu} \\
        &\qquad + (\mu \rightarrow -\mu)\Bigg\}.
        \end{aligned}
     \end{equation}
Replacing $\epsilon(k+q/2) = Tx -\mu/2 + {\bf vq}/2$, we obtain
    \begin{eqnarray}
    && \chi(q) = \label{chiP}\\
    &&\frac{3\rho T}{2}\int \frac{d\Omega}{4\pi}\int_{-D/2T}^{D/2T} \frac{dx}{x} \frac{\sinh 2x}{\cosh^2 x +\sinh^2(\mu +{\bf vq})/4T}. \nonumber
    \end{eqnarray}
The function $A(\mu/4T,\, qv/4T) \equiv 1 - \chi(q;T,\mu)/\chi(0,T,\mu)$ is plotted in Fig.~\ref{fig:Stiff3} for representative values of $\mu/4T$.

\begin{figure}[t]
            \centering
            \includegraphics[width=0.8\linewidth]{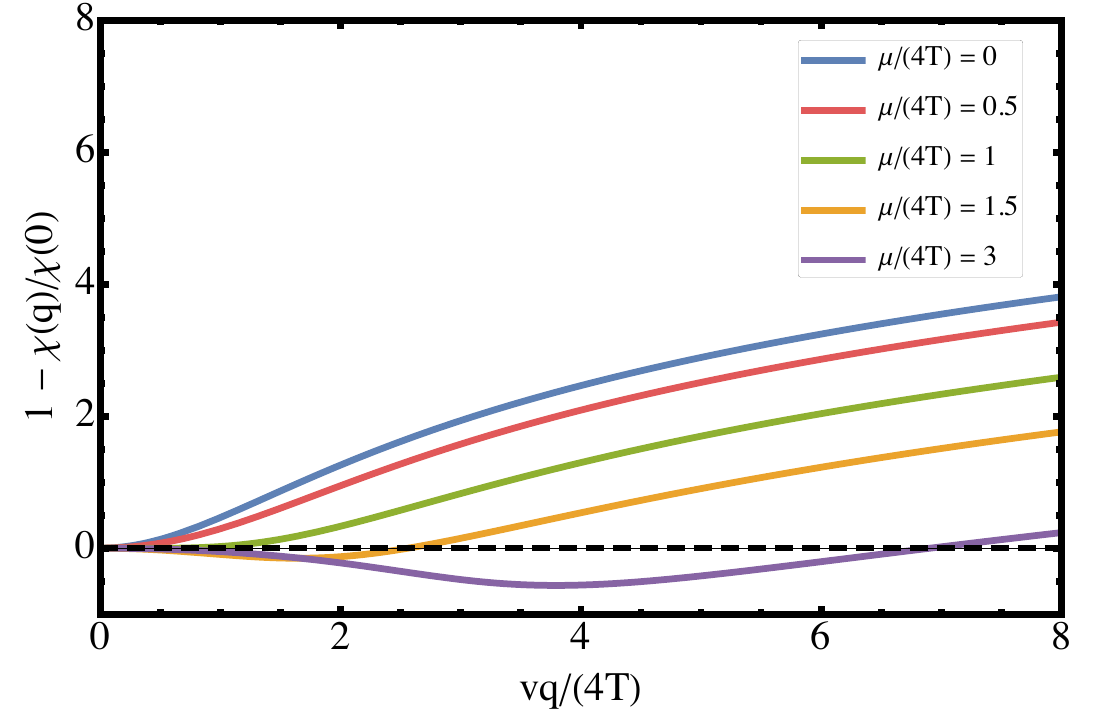}
            \caption{Function $A(\mu/4T, vq/4T)$ for $\mu/4T = 0$ (light blue), $0.5$ (yellow), $1$ (green), $1.5$ (rose), $3$ (purple).
            }
            \label{fig:Stiff3}
        \end{figure}

Expanding Eq.~(\ref{chiP}) at small $q$ gives
\begin{eqnarray}
&& A(\mu,q) \approx \label{stiff}\\
&&\frac{\langle(\vec v\vec q)^2\rangle}{ T^2}f(\mu/4T) + \frac{\langle(\vec v\vec q)^4\rangle}{ (4T)^4}g(\mu/4T),\nonumber \\
&&f(y) = \frac{1}{16}\int_0^{\infty}\frac{d x}{x}\frac{\sinh(2x)}{(\cosh^2x +\sinh^2y)^2}\times\nonumber\\
&&\Big[\cosh(2y) - \frac{\sinh^2(2y)}{\cosh^2 x +\sinh^2y}\Big]\nonumber.
\end{eqnarray}
Due to cubic symmetry, the leading correction depends on $q^2 = q_x^2+q_y^2+q_z^2$. Neglecting cubic anisotropy ($|\mathbf{v}| = \mathrm{const}$), the functions $f(y)$ and $g(y)$ are shown in Fig.~\ref{fig:Stiff}.

\begin{figure}[!htbp]
    \centering
    \begin{minipage}[t]{0.7\linewidth}
        \includegraphics[width=\linewidth]{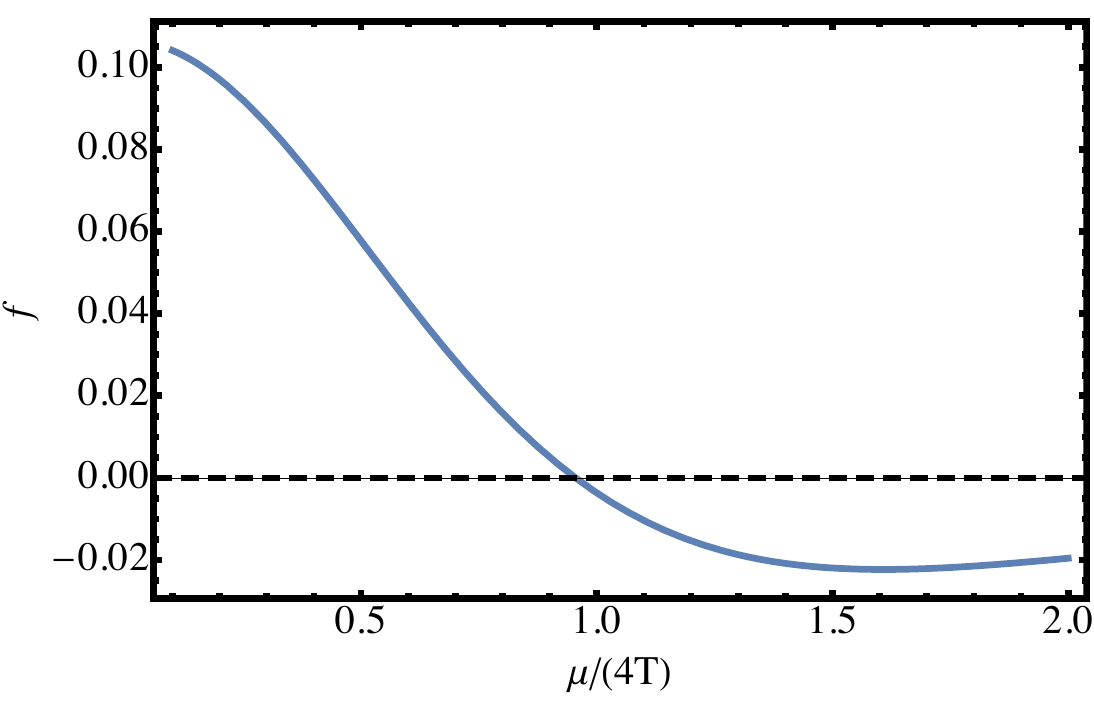}
    \end{minipage}\hfill
    \begin{minipage}[t]{0.7\linewidth}
        \includegraphics[width=\linewidth]{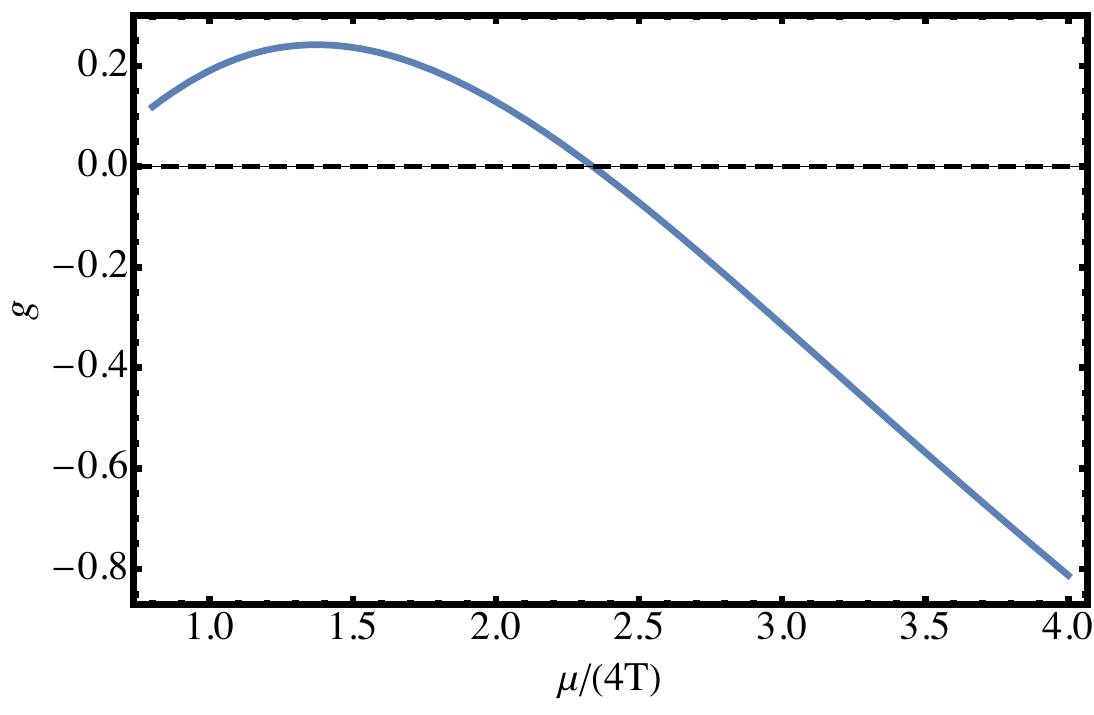}
    \end{minipage}
    \caption{(a) Function $f(\mu/4T)$ and (b) function $g(\mu/4T)$, entering Eq.~(\ref{stiff}).}
    \label{fig:Stiff}
\end{figure}

These microscopic results feed directly into the phase diagram discussed in the main text. In particular, they identify two characteristic temperatures: one, $T_{c1}$, at which $\tau(T_{c1},\mu) = 0$, and another, $T_{c2}$, at which $f(T_{c2},\mu) = 0$. If $T_{c2} > T_{c1}$, the mean-field transition proceeds directly to the IC phase (right-hand side of Fig.~\ref{fig:Phase_D}). Otherwise, the system first enters the commensurate phase, and the pairing susceptibility must be recalculated to determine whether a further transition occurs at lower temperature.

\begin{figure}[!htbp]
            \centering
            \includegraphics[width=0.9\linewidth]{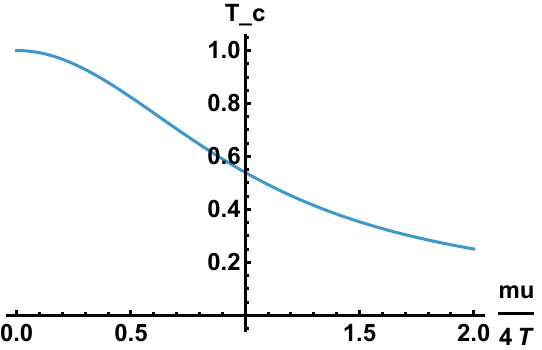}
            \caption{Mean-field $T_c/T_{c0}$ vs.\ $\mu/4T$, Eq.~(\ref{Tc}), for $K=t$. The vertical line at $\mu/4T = 1$ marks where the order-parameter stiffness vanishes.
            }
            \label{fig:Phase_Diagram}
        \end{figure}

For $K = t$, the calculations show that the quadratic stiffness coefficient vanishes ($f = 0$) at $\mu \approx 4T$. At the same time,
 \begin{eqnarray}
 && 2\tau =\label{Tc}\\
 && \ln(T/T_{c0}) + \int_0^{\infty} d x\frac{\tanh x}{x}\frac{\sinh^2(\mu/4T)}{\cosh^2 x + \sinh^2(\mu/4T)}.\nonumber
 \end{eqnarray}

At $\mu = 4T$ we have $\tau \approx \ln(T/T_{c0}) + 0.617$, meaning that the mean-field transition temperature drops by approximately a factor of $1.85$. However, as the system approaches the curve $T = \mu/4$, increasing fluctuations drive the transition temperature further down, as illustrated in Fig.~\ref{fig:Phase_Diagram}.

\balance
\bibliographystyle{apsrev4-2}
\bibliography{PDWinCPT_X.bib}
    \end{document}